\documentclass[preprint,12pt]{elsarticle}
\usepackage[left=1in, right=1in, top=1in, bottom=1.25in]{geometry}
\usepackage{amssymb}
\usepackage{soul}
\usepackage{amsthm}
\usepackage{xtab,afterpage}
\usepackage{lipsum}
\usepackage{amsmath}
\usepackage{soul}
\usepackage{subfig}
\usepackage{multirow}
\usepackage{graphicx}
\usepackage{dcolumn}
\usepackage{xcolor}
\usepackage{physics}
\usepackage{graphicx, float}
\usepackage{amsmath}
\usepackage{tikz}
\usepackage{hyperref}
\usepackage{physics}
\usepackage{amsmath}
\usepackage{mathdots}
\usepackage{yhmath}
\usepackage{cancel}
\usepackage{color}
\usepackage{siunitx}
\usepackage{array}
\usepackage{multirow}
\usepackage{amssymb}
\usepackage{gensymb}
\usepackage{tabularx}
\usepackage{booktabs}
\usepackage{xcolor}
\usepackage{ulem}
\usepackage{nicematrix}
\usepackage{arydshln}
\usepackage{bm}
\usepackage{hyperref}
\usepackage{setspace}

\journal{}

\begin{document}
\begin{frontmatter}

\title{Machine Learning-Driven Insights into Excitonic Effects in 2D Materials}
\author{Ahsan Javed$^{1,2}$, Sajid Ali$^{\ast,3}$}
\address{$^1$Department of Physics, Syed Babar Ali School of Science and Engineering, Lahore University of Management Sciences (LUMS), Lahore-54792, Pakistan}
\address{$^2$Department of Physics, COMSATS University Islamabad, Lahore Campus, Lahore, Pakistan}
\address{$^3$Department of Materials Science and Engineering, Monash University, Victoria 3800, Australia}

\address{$^{\ast}$Corresponding Author : sajid.ali@monash.edu}
\date{\today}
\begin{abstract}
Understanding excitonic effects in two-dimensional (2D) materials is critical for advancing their potential in next-generation electronic and photonic devices. In this study, we introduce a machine learning (ML)-based framework to predict exciton binding energies in 2D materials, offering a computationally efficient alternative to traditional methods such as many-body perturbation theory (GW) and the Bethe-Salpeter equation. Leveraging data from the Computational 2D Materials Database (C2DB), our ML models establish connections between cheaply available material descriptors and complex excitonic properties, significantly accelerating the screening process for materials with pronounced excitonic effects. Additionally, Bayesian optimization with Gaussian process regression was employed to efficiently filter materials with largest exciton binding energies, further enhancing the discovery process. Although developed for 2D systems, this approach is versatile and can be extended to three-dimensional materials, broadening its applicability in materials discovery.
\end{abstract}
\end{frontmatter}
\section{Introduction}
\noindent Two-dimensional materials (2DMs) have garnered substantial interest due to their exceptional optical and electronic properties, positioning them as promising candidates for next-generation optoelectronic technologies. A defining feature of these materials is their pronounced excitonic effects, which are significantly more pronounced in 2D systems compared to bulk materials \cite{qiu2013optical, chernikov2014exciton}. This enhanced excitonic behavior arises from reduced dielectric screening and increased Coulomb interactions in monolayers \cite{reviewTMDCS_2015, Ashwin_2012}. Consequently, understanding these excitonic properties is crucial for optimizing the performance of devices based on such materials. \\

\noindent The exciton binding energy (EBE) is a fundamental property of excitons, defined as the energy required to dissociate a bound electron-hole pair (exciton) into free charge carriers. In first-principles calculations, EBE determination typically requires computationally intensive methods such as the GW approximation and then Bethe-Salpeter equation (BSE) \cite{javed2024investigation}. The EBE is then calculated as:
\begin{equation}
    EBE = E_g^{electronic} - E_g^{optical}
\end{equation}
Here $E_g^{electronic}$ is the G$_o$W$_o$-corrected quasi-particle band gap, $E_g^{optical}$ is energy of the lowest-energy bright exciton from BSE \cite{qiu2017}.
The GW approximation surpasses mean-field, independent-particle DFT by accounting for many-body electron-electron interactions, offering a more accurate understanding of electronic properties, including excitation energies, band gaps, and optical characteristics. This method involves iteratively solving for the Green's function ($\mathbf{G}$), the screened Coulomb interaction ($\mathbf{W}$), and the self-energy ($\mathbf{\Sigma}$) until self-consistency is achieved, yielding an improved description of the electronic structure \cite{GW_BSE}. Following the GW calculation, the BSE method is employed to compute excitonic effects by incorporating electron-hole interactions, providing precise estimates of EBE through electron-hole correlation \cite{Ashwin_2012}. However, these methods are computationally demanding, making them less practical for large-scale material screening. \\

\noindent To address this challenge, we propose a machine learning-assisted approach for efficiently predicting EBE. This method leverages data such as band gaps obtained from PBE and structural parameters available in the C2DB database. By correlating these simpler features with exciton binding energy estimates, which would typically require advanced computational methods, this approach accelerates the materials discovery process. \\

\noindent While Liang et al. (2019) pioneered physics-inspired machine learning for band gap and EBE prediction based on Phillips ionicity theory, their reliance on small datasets and ad-hoc structural maps limits the generalizability of their approach. Additionally, their model only covers A$_m$B$_n$ materials, whereas the C2DB has been significantly updated over the past five years \cite{liang2019phillips}. In another study, Lin et al. (2023) used features such as the highest occupied molecular orbital (HOMO) and lowest unoccupied molecular orbital (LUMO) from the C2DB database to predict EBE in 2D materials \cite{lin2023machine}. Their model achieved a $R^2$ value of 0.80 and a mean absolute error (MAE) of 0.21 eV with a gradient boosting (GB) regression model. Other models have the potential to improve this accuracy, as we will demonstrate in this work. While machine learning has been employed to predict EBE, Bayesian Optimization remains unexplored in this context. This combination of scale, methodology, and accuracy marks a substantive advance over prior ML-based EBE studies. \\

\noindent Machine learning (ML) approaches require a fine balance between accuracy and computational efficiency when predicting complex quantities from simpler, readily available parameters \cite{sajid2022spin}. This study highlights the potential of ML to streamline the identification of 2D materials with large excitonic effects, enabling faster screening for optoelectronic applications.

\section{Methodology}
\subsection{Dataset}
In the context of machine learning in materials science, material databases are essential for the success of predictive models. Using a large amount of high-quality data is mandatory to achieve a robust and accurate predictions. In this study, we used C2DB database \cite{c2db2018, C2DB_2021} to train and evaluate our machine learning models, which comprises band gap data for transition metal dichalcogenides (e.g., MoS$_2$) \cite{band}, transition metal oxides and other technologically important materials like hexagonal boron nitride \cite{sajid2020single}, MgI$_2$, MgBr$_2$ \cite{ali2023high} etc. \\ 

\noindent In our study, the data is randomly partitioned into two sets: a training dataset and a test dataset. The partitioning allocates 60\% of the data for training and 40\% for testing, a ratio deemed optimal for achieving accurate machine learning predictions \cite{optimaljoseph2022}.

\subsection{Features selection}
Pearson correlation coefficient quantifies the linear relationship between variables, providing insight into how one property influences another in monolayer materials. In this study, it serves as a critical tool to evaluate the dependencies between different features \cite{wei2019machine, zhang2018strategy}. A high Pearson coefficient indicates a strong linear association, such as between PBE and HSE06 band gaps, suggesting that trends in simpler PBE calculations can predict more computationally intensive HSE06 results. Similarly, correlations between G$_0$W$_0$ band gaps and other properties reflect how electronic interactions evolve between computational methods. The wrapper method evaluates the predictive power of features by assessing their impact on model performance. Unlike statistical filter methods, it iteratively trains a model with different feature subsets, capturing both linear and nonlinear dependencies \cite{zhong2022explainable_wrapper}. Understanding these relationships helps optimize feature selection for predicting EBE. Based on these methods, the features selected for this study include the layer thickness, layer group number, atomic number, valence electron count, atomic number, ionic radii difference, and PBE band gap of the monolayer.
\subsection{Algorithm selection}
To predict the exciton binding energy, we evaluated multiple machine learning algorithms, each with distinct advantages. Neural Networks (NNs) excel at capturing complex, nonlinear relationships by adjusting weights through layers of interconnected neurons \cite{kang2015machine}. Random Forest (RF) and Gradient Boosting (GB) are ensemble tree-based methods \cite{nadkarni2023comparative}: RF combines predictions from multiple decision trees to reduce variance, while GB sequentially builds trees to minimize prediction error. Support Vector Machines (SVM) define optimal hyperplanes to separate data, useful for high-dimensional spaces \cite{salcedo2014support}. Kernel Ridge Regression (KRR) combines ridge regularization with kernel methods to handle nonlinearity \cite{vovk2013kernel}. We used these models as implemented in scikit-learn \cite{scikit-learn} package. In the following section, we present the results of these ML models in predicting EBE.. 
\subsection{Workflow}
The workflow of this study proceeds in three stages. First, the dataset is constructed using structural and electronic descriptors obtained from the C2DB database. Second, multiple machine learning models are trained and evaluated. Finally, both the G$_0$W$_0$ quasiparticle band gaps and the exciton binding energies are predicted directly from the selected descriptors using the trained models.
\section{Results and discussion}
\subsection{Predicting Quasi-particle energies}
\begin{figure}[h!]
    \centering
    \includegraphics[scale=0.95]{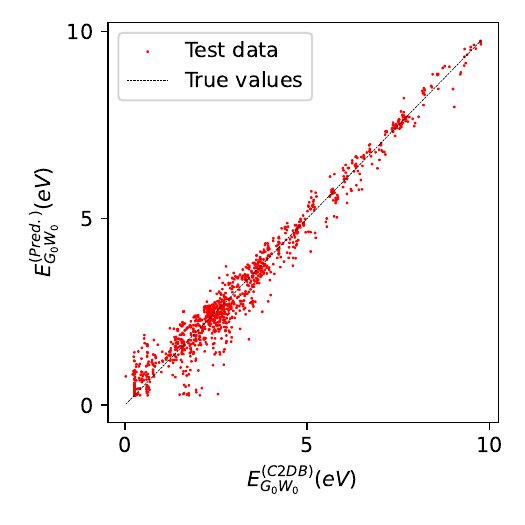}
    \caption{Random forest-based machine learning model for predicting G$_0$W$_0$ band gaps in 2D materials.}
    \label{fig:first}
\end{figure}

\begin{figure}[h!]
    \centering
    \includegraphics[scale=0.9]{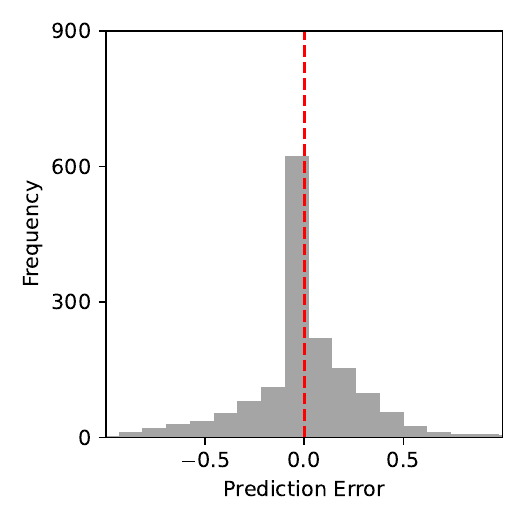}
    \caption{Distribution of prediction errors for G$_0$W$_0$ band gaps using the Random Forest model, showing a peak at zero, indicating high accuracy with minimal deviation between predicted and actual values.}
    \label{fig:rf}
\end{figure}

The Random Forest regression model  demonstrates strong performance in predicting G$_0$W$_0$ band gaps, which correspond to QP energies. The model's effectiveness is illustrated in Figure \ref{fig:first}. The distribution of prediction errors for the G$_0$W$_0$ band gaps demonstrates a strong alignment with actual values from the C2DB database, indicating the model’s high predictive accuracy and reliability. The histogram of errors has a clear peak around zero, as shown in Figure \ref{fig:rf}, indicating that most of the predictions are very close to the actual band gap values, R$^2$ equals to 0.98, with MAE and RMSE values of 0.20 and 0.33, respectively.   
\subsection{Exciton binding energy}
\begin{figure}[h!]
    \centering
    \includegraphics[scale=1]{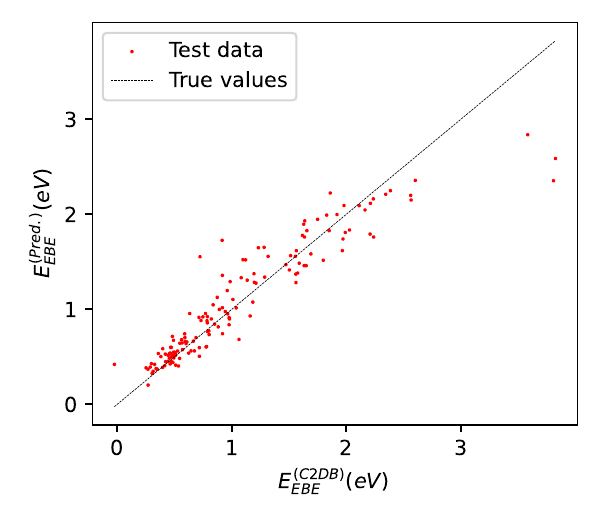}
    \caption{Exciton Binding Energy predicted using the RF model, illustrating the relationship between machine learning predictions and C2DB values, with a focus on capturing the accuracy and reliability of excitonic effects in 2D materials.}
    \label{fig:ebe}
\end{figure}

\begin{figure}[h!]
    \centering
    \includegraphics[scale=1]{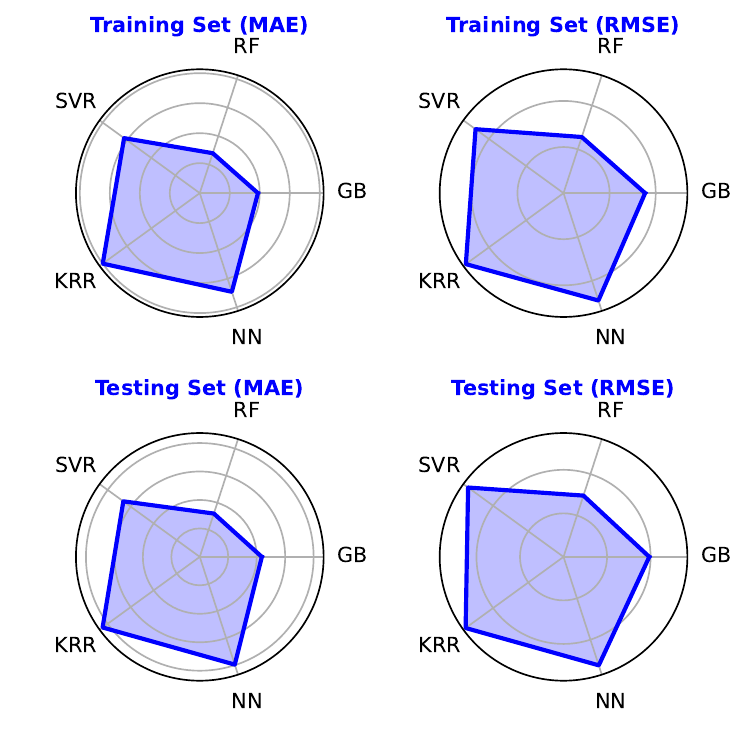}
    \caption{Radar plots comparing the performance of different ML algorithms in predicting exciton binding energy. The area under each curve represents the MAE and RMSE, with larger areas indicating higher error. The RF model shows the smallest area, indicating the lowest prediction error.}
    \label{fig:radar}
\end{figure}

The exciton binding energy is the energy required to separate an exciton into free, independent charge carriers. The EBE prediction model, trained using the RF algorithm, is presented in Figure \ref{fig:ebe}. This model effectively captures the EBE with high accuracy for the available data in the C2DB database, achieving an R$^2$ value of 0.84.   
 \\
The radar plots in Figure \ref{fig:radar} illustrate the performance of five ML algorithms i.e. Gradient Boosting, Random Forest, Support Vector Regression, Kernel Ridge Regression, and Neural Networks, across training and testing sets, evaluated by mean absolute error (MAE), root mean square error (RMSE), and R$^2$ metrics. \\

\noindent These metrics collectively assess each model’s prediction accuracy and capacity to generalize to new data set. In both training and testing sets, the RF model demonstrates superior performance, achieving a training MAE of 0.070 and RMSE of 0.126. Its testing metrics are also impressive, with MAE of 0.081 and RMSE of 0.148, indicating strong generalization with a minimal increase in error on unseen data. R$^2$ scores of 0.84 signifies that it captures most of the variance in the exciton binding energy as shown in Figure \ref{fig:radar}. \\
 \\
Gradient Boosting model follows closely, with an R$^2$ of 0.80 as well in other statistical parameters as illustrated in Fig. \ref{fig:radar}. The Neural Network, KRR, and SVR algorithms relatively under-perform having R$^2$ of around 0.56. Random Forest is particularly effective in modeling exciton binding energy due to its ensemble nature, which combines multiple decision trees to reduce over fitting and improve generalization. It captures complex, nonlinear relationships between the features and the target variable with high accuracy. RF also handles feature interactions well and is robust to noise, making it suitable for datasets with moderate complexity, as demonstrated by its consistently low MAE and RMSE on both training and test sets. \\

While Lin et al. \cite{lin2023machine} employed HOMO and LUMO energies from the C2DB database as primary features for predicting EBE and achieving R$^2$ of 0.80 using GB algorithm. Our Random Forest model demonstrates superior predictive performance through incorporation of additional physically meaningful descriptors and yielding improved accuracy with R$^2$ equal to 0.84. This advancement highlights the importance of comprehensive feature engineering in excitonic property prediction.
\subsection{SHAP Analysis}
To interpret the non-linear predictions of exciton binding energy made by the RF model, we employed SHAP (SHapley Additive exPlanations), a game-theoretic approach that assigns each feature an importance value for individual predictions. SHAP values quantify how much each descriptor increases or decreases the predicted EBE, allowing both a global ranking of feature importance and local insight into directionality \cite{SHAP}. Unlike linear regression coefficients, SHAP is model-agnostic and captures the complex non-linear dependencies present in ensemble methods. \\

\begin{figure}[h!]
    \centering
    \includegraphics[scale=0.8]{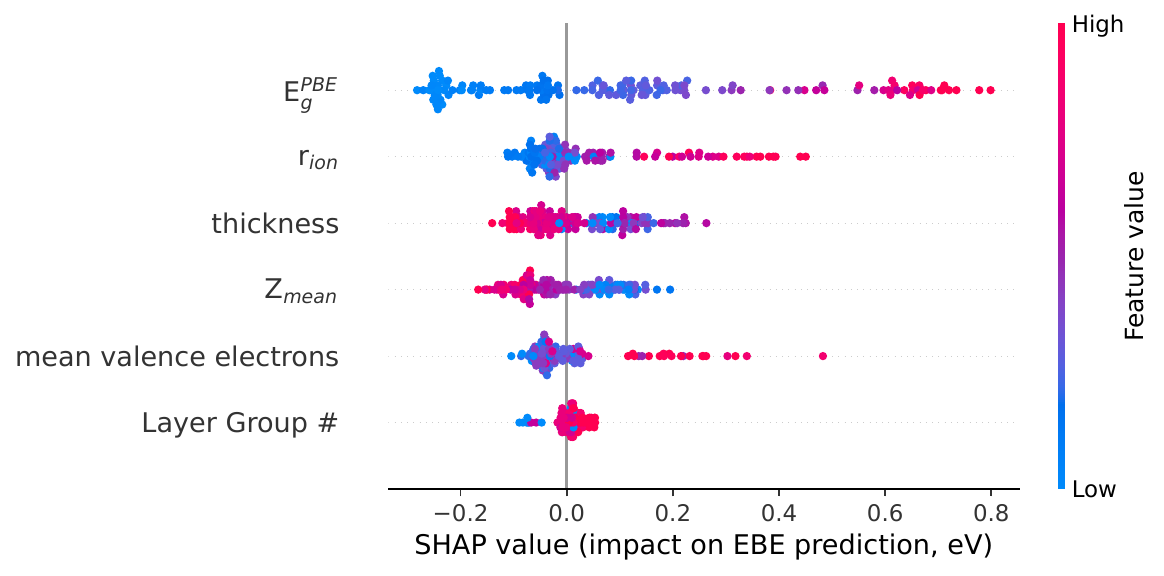}
    \caption{SHAP (SHapley Additive exPlanations) summary plot for the RF model predicting EBE.}
    \label{fig:shap}
\end{figure}

Figure \ref{fig:shap} presents the SHAP summary plot for the top features in our model. The PBE band gap emerges as the dominant descriptor, with larger band gaps strongly increasing predicted EBE, consistent with reduced dielectric screening in wide-gap systems. Layer thickness also makes a significant contribution, as it correlates with quantum confinement and dielectric screening, both of which strongly influence exciton binding. The mean atomic number (Z$_{mean}$) ionic radius, and layer group number show more modest but still notable effects, shaping the electronic environment and excitonic interactions.
\subsection{Bayesian Optimization}
This study implements a Bayesian optimization (BO) framework to identify 2D materials with highest excitonic binding energy from the C2DB database. BO is a data-driven, efficient approach that combines surrogate modeling with iterative optimization to explore the dataset and predict materials with desired properties. The framework leverages Gaussian Process Regression (GPR) as the surrogate model, which not only predicts EBE but also estimates uncertainties, enabling a balance between exploration of uncertain regions and exploitation of high-performing candidates \cite{snoek2012practical}. \\ 

The GPR model employs a composite kernel comprising a radial basis function, a dot-product kernel, and a constant kernel. This kernel effectively captures complex non-linear relationships in the dataset while ensuring numerical stability during optimization. The hyperparameters of the kernel are optimized during each iteration using the L-BFGS-B algorithm to ensure accurate surrogate modeling \cite{di2016optimization}. This BO algorithm explores a search space of 4001 two-dimensional materials from the C2DB database. 
\begin{figure}[h!]
    \centering
    \includegraphics[scale=0.35]{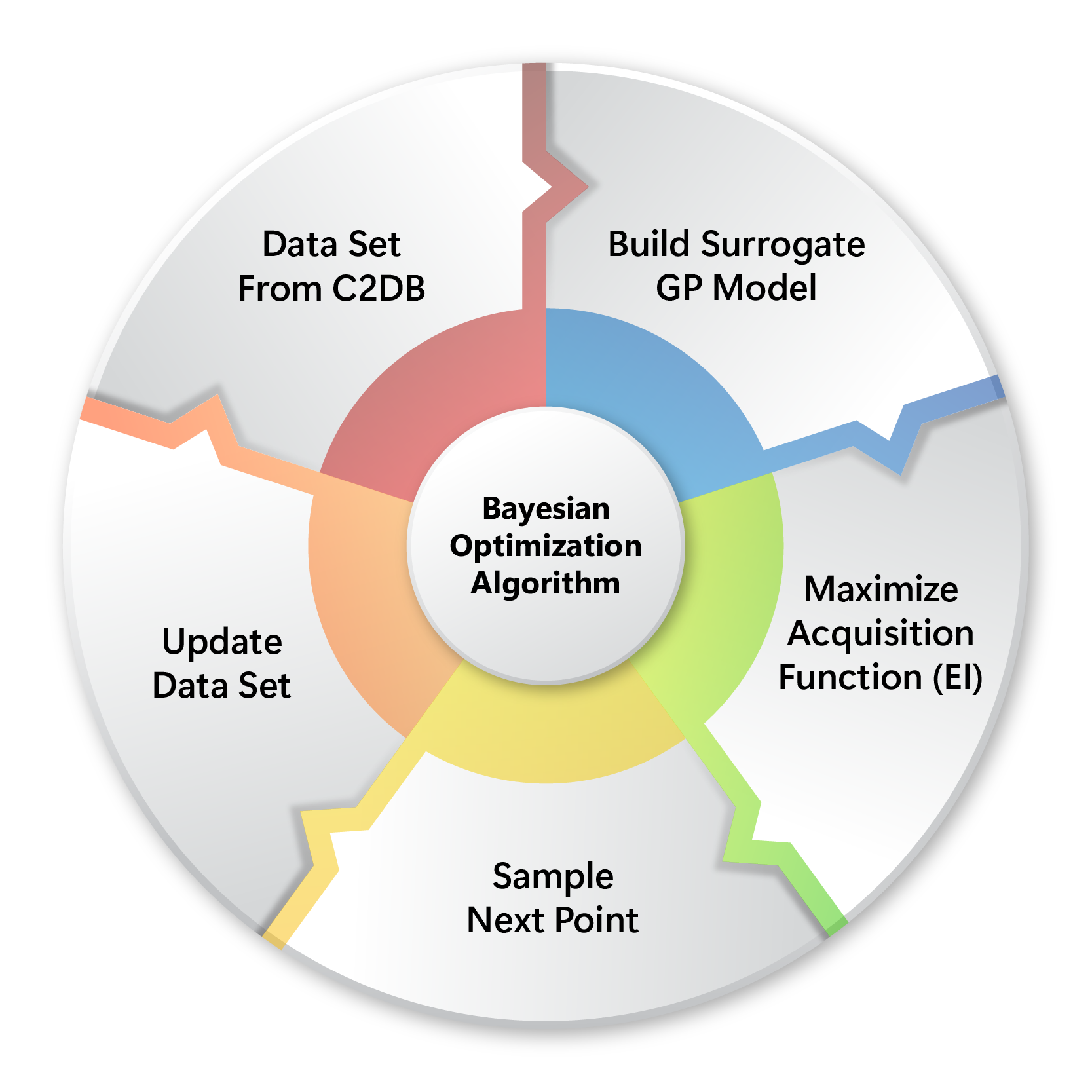}
    \caption{A Bayesian optimization algorithm is implemented, starting with a dataset from C2DB containing 4001 monolayers. Gaussian Process Regression is used as the surrogate model, employing Expected Improvement as the acquisition function, with the dataset updated iteratively.}
    \label{fig:bo}
\end{figure} \\
An important function is Expected Improvement (EI) acquisition function, which directs the search by selecting materials expected to offer the greatest improvement in EBE. The optimization workflow involves iteratively training the GPR model on available data, computing EI for all candidates, and selecting the material with the highest EI for evaluation \cite{frazier2018bayesian}. The predicted EBE is then added to the training dataset, and the process repeats until convergence or a predefined number of iterations is reached. The EI used in the current study is given by:
\[
\text{EI} = (\mu - f_{\text{best}}) \cdot \Phi(Z) + \sigma \cdot \phi(Z)
\]

\noindent Where the standardized improvement \( Z \) is expressed as:
\[
Z = \frac{\mu - f_{\text{best}}}{\sigma}
\]
Here $\mu$ is the predicted mean, $\sigma$ is the predicted uncertainty, $f_{\text{best}}$ is the current best EBE, $\Phi$ and $\phi$ are the cumulative and probability density functions of the standard normal distribution respectively. This BO-based approach is particularly advantageous for predicting EBE in 2D materials, where experimental or computational evaluations are resource-intensive. By prioritizing candidates with the highest potential, the framework reduces the number of evaluations required while ensuring that top-performing materials are identified. The results highlight Bayesian optimization as a powerful and scalable methodology for accelerating the discovery of 2D materials with desirable excitonic properties. \\

\noindent The table \ref{tab:monolayers} lists the highest predicted EBE values through BO. Additionally, our model has also predicted EBE values for monolayers that are not covered in the C2DB database. \\
\begin{table}[h!]
    \centering
    \caption{Monolayers with highest exciton binding energies}
    \begin{tabular}{@{}lccc@{}}
    \hline \hline
        \textbf{Formula} & \textbf{Space Group} & \textbf{EBE (Predicted)} & \textbf{EBE (C2DB)} \\ \hline 
        Li\(_3\)Cl\(_3\) & p4/mmm & 2.67 & \\
        SrCl\(_2\) & p-6m2 & 2.63 & 2.63 \\
        Y\(_2\)F\(_2\)O\(_2\) & p-3m1 & 2.62 & \\
        CaCl\(_2\) & p-6m2 & 2.61 & 2.60 \\
        Mg\(_4\)Cl\(_8\) & pmmm & 2.60 & \\
        Be\(_2\)Br\(_2\)Cl\(_2\) & pm2\_1b & 2.60 & \\
        BaCl\(_2\) & p-6m2 & 2.60 & 2.58 \\   
        Li\(_6\)Cl\(_6\) & p-3m1 & 2.60 & \\
        Ca\(_2\)Cl\(_4\)H\(_8\)O\(_4\) & pman & 2.59 & \\
        CaCl\(_2\) & p-3m1 & 2.58 & 2.56 \\
        \hline \hline
    \end{tabular}
    \label{tab:monolayers}
\end{table} \\
The high EBE observed in monolayers of alkaline earth metal chlorides i.e., ACl$_2$ where A = Sr, Ca, Ba and other 2D materials containing chlorine. The atoms of chlorine have a high electron affinity, which contributes to stronger Coulombic attraction between electrons and holes. This enhances the binding energy of excitons, particularly in low-dimensional systems where dielectric screening is already reduced. \\

\noindent The table \ref{tab:2H} tabulates the predicted EBE for TMDC monolayers, which are well-known for their excitonic properties. The strong excitonic effects are attributed to their direct band-gap in monolayer form, which enhances their optical absorption capabilities. \\
\begin{table}[h!]
    \centering
    \caption{TMDC monolayers with highest exciton binding energies}
    \begin{tabular}{@{}lcc@{}}
    \hline \hline
        \textbf{Formula} & \textbf{EBE (Predicted)} & \textbf{EBE (C2DB)} \\ \hline 
        HfS\(_2\) & 1.15 & 1.29 \\
        ZrS\(_2\) & 1.14 & 1.18  \\
        HfSe\(_2\) & 1.14 & 0.95 \\
        TiS\(_2\) & 0.94 & 0.98 \\ 
        ZrSe\(_2\) & 0.89 & 0.90 \\ 
        .. & .. & .. \\
        MoS\(_2\) &  0.52 & 0.55 \\
        WS\(_2\) & 0.53 & 0.52 \\
        WTe\(_2\) & 0.46 & 0.42 \\
        \hline \hline
    \end{tabular}
    \label{tab:2H}
\end{table}\\
MoS$_2$, the most studied TMDC monolayer, is reported to exhibit an exciton binding energy in the range of 0.5–0.8 eV \cite{C2DB_2021,saigal2016exciton}, which is consistent with our model’s predicted value (see Table 2). This agreement with literature benchmarks validates the reliability of our approach. At the same time, Table 2 highlights several other monolayers with predicted EBEs comparable to or exceeding that of MoS$_2$, suggesting promising yet underexplored candidates for future investigations. 
\section{Conclusion}
In this study, we demonstrated a machine learning-assisted approach for predicting exciton binding energies in two-dimensional materials, using band gap data from simple DFT calculations. By training and evaluating multiple ML algorithms, we found that the RF model provided the most reliable predictions, effectively bridging the gap between computational efficiency and accuracy. Our model offers a rapid and cost-effective alternative to traditional GW and BSE methods, enabling faster screening and discovery of materials with significant excitonic effects. Additionally, we implemented a Bayesian optimization framework, which further streamlined the identification of top EBE monolayers by efficiently guiding the search for promising candidates.  This integration of BO with machine learning underscores the potential for data-driven approaches to revolutionize materials discovery processes.

\section{CRediT author statement}
\textbf{A. Javed:} Conceptualization, Methodology, Software, Writing - Original Draft \textbf{A. Sajid:} Supervision, Validation, Writing - Review \& Editing.  

\section{Competing interests}
The authors declare no competing interests.

\bibliographystyle{elsarticle-num-names}
\bibliography{reference}
\end{document}